# Modeling Oral Multispecies Biofilm Recovery After Antibacterial Treatment

Xiaobo Jing[1#], Xiangya Huang[2,3#], Markus Haapasalo[3], Ya Shen[3*] and Qi Wang[1,4,5*]

[1]Beijing Computational Science Research Center, Beijing 100193, China;

[2]Department of Conservative Dentistry and Endodontics, Guanghua School of Stomatology, Guangdong Province Key Laboratory of Stomatology, Sun Yat-Sen University, 510055 Guangzhou, Guangdong, China;

[3]Division of Endodontics, Department of Oral Biological and Medical Sciences, Faculty of Dentistry, University of British Columbia, Vancouver, British Columbia, V6T 1Z3 Canada;

[4]Department of Mathematics, University of South Carolina, Columbia, SC 29208, USA;

[5]School of Materials Science and Engineering, Nankai University, Tianjin 300071, China.

[#]These authors contributed equally to this work.

**Corresponding author:** Prof. Qi Wang (email: qwang@math.sc.edu)
Prof. Ya Shen (email: yashen@dentistry.ubc.ca)

## Author Contributions

Xiaobo Jing developed the model, carried out the simulations and wrote the modeling portion of the manuscript; Xiangya Huang conducted the experiment and analyzed the experimental data; Markus Haapasalo designed the experiment, was involved in data analysis and writing of the manuscript; Ya Shen designed the experiment, analyzed the data and wrote the experimental portion of the manuscript; Qi Wang developed the model, analyzed the numerical results, and wrote the manuscript. All the authors reviewed the manuscript.

## Competing interests
The authors declare no competing interests.

## Abstract

Recovery of multispecies oral biofilms is investigated following treatment by chlorhexidine gluconate (CHX), iodine-potassium iodide (IPI) and Sodium hypochlorite (NaOCl) both experimentally and theoretically. Experimentally, biofilms taken from two donors were exposed to the three antibacterial solutions (irrigants) for 10 minutes, respectively. We observe that (a) live bacterial cell ratios decline for a week after the exposure and the trend reverses beyond a week; after fifteen weeks, live bacterial cell ratios in biofilms fully return to their pretreatment levels; (b) NaOCl is shown as the strongest antibacterial agent for the oral biofilms; (c) multispecies oral biofilms from different donors showed no difference in their susceptibility to all the bacterial solutions. Guided by the experiment, a mathematical model for biofilm dynamics is developed, accounting for multiple bacterial phenotypes, quorum sensing, and growth factor proteins, to describe the nonlinear time evolutionary behavior of the biofilms. The model captures time evolutionary dynamics of biofilms before and after antibacterial treatment very well. It reveals the crucial role played by quorum sensing molecules and growth factors in biofilm recovery and verifies that the source of biofilms has a minimal to their recovery. The model is also applied to describe the state of biofilms of various ages treated by CHX, IPI and NaOCl, taken from different donors. Good agreement with experimental data predicted by the model is obtained as well, confirming its applicability to modeling biofilm dynamics in general.

# Introduction

Success in endodontics is achieved by a combination of host and treatment factors that contribute to the management of the infection, prevention and healing of periapical pathosis [1,2]. Instrumentation is important in removing microbes from the root canals, but after mechanical instrumentation alone, canals often remain heavily infected. Instrumentation creates the space necessary for effective irrigation, which plays the key role in further reducing the number of residual microbes. Irrigating solutions with a strong antibacterial effect are necessary. However, currently available irrigants face great challenges in eliminating all the biofilms from the root canals. The biofilms are results of the microbial growth, where dynamic communities of interacting sessile cells are irreversibly attached to a solid substratum as well as next to each other through a network of extracellular polymeric substances (EPS). Microbial communities growing in biofilms are very difficult to be eradicated with antibacterial agents [3,4]. Microorganisms in mature biofilms can be extremely resistant to antibacterial agents for reasons that have yet been fully understood [5,6].

There are several antibacterial solutions or irrigants available in the market currently. Sodium hypochlorite (NaOCl) is the most popular and important irrigating solution [7]. In water, NaOCl ionizes into sodium ion, Na+, and hypochlorite ion, OCl-, establishing equilibrium with hypochlorous acid (HOCl). Hypochlorous acid is responsible for its antibacterial activity; OCl- is less effective than undissolved HOCl. NaOCl is commonly used in concentrations ranging from 0.5% to 6%. It is the only irrigant in endodontics that can dissolve organic tissue, including the organic part of the smear layer. Another irrigant, Chlorhexidine digluconate (CHX), is widely used in disinfection in dentistry because of its antibacterial activity [8-10]. It has gained considerable popularity in endodontics as an irrigating solution and as an intra-canal medicament. However, CHX has no tissue dissolving capability and therefore it cannot replace sodium hypochlorite. CHX is membrane permeable and attacks the bacterial cytoplasmic inner membrane or the yeast plasma membrane. In high concentrations, CHX causes coagulation of intracellular components [11]. One of the reasons for the popularity of CHX is its substantivity (i.e. continued antibacterial effect), which stems from its ability to bind to hard tissue and maintain antibacterial activity. Iodine compounds are among the oldest disinfectants still actively used. They are best known for their use on surfaces, skin, and operation fields. Iodine is less reactive than the chlorine in hypochlorite. However, it kills rapidly and has bactericidal, fungicidal, tuberculocidal, virucidal, and even sporicidal activity [12].

Iodine penetrates rapidly into the microorganisms and causes cell death by attacking proteins, nucleotides, and other key subcellular components of the cell [12,13].

Oral biofilm recovery after treatment by two different CHX irrigants for 1, 3, 10, minutes respectively was recently studied by Shen et al [14]. Results from the study showed that biofilms started recovering after two weeks, and then fully returned to the pre-treatment level after eight weeks. However, because all the biofilms in that study were grown from plaque bacteria obtained from one donor and only the CHX solution was tested, it is insufficient to evaluate whether the results can be generalized or if they represent the behavior of that particular biofilm exposed to CHX only. Therefore, the experimental aim of this study is to assess the effect of the source of biofilm bacteria and the type of antibacterial agents on biofilm recovery after exposure to different antibacterial agents.

In this paper, we investigate recovery of multispecies oral biofilms from two donors following CHX, iodine-potassium iodide (IPI) and sodium hypochlorite (NaOCl) treatments, respectively. Before the antibacterial treatment, the biofilms are allowed to grow for three weeks. After 10 minutes' treatment by the antibacterial agents, a large number of bacteria in the biofilms continue to die up to one week and the bacterial dying rate reduces beyond one week. The full recovery of the biofilms needs about 15 weeks afterwards. Interestingly, recovery of biofilms from different donors shows similar dynamics after the treatment.

To fully understand the mechanism of biofilm recovery, a quantitative model in the form of dynamical systems is developed, which accounts for the regulation of growth factor proteins and quorum sensing molecules to bacterial growth and EPS regulation in addition to bacterial cell interactions and their response to antibacterial treatment. The model captures time evolutionary dynamics of the biofilms after antibacterial treatment. The model is then applied to another experiment on biofilm's susceptibility to three antibacterial agents right after antibacterial treatment to confirm its validity in modeling biofilm dynamics for biofilms of various ages right after the treatment [6]. Using the model, we can explore crucial roles played by the EPS, quorum sensing molecules and growth factor proteins on biofilm dynamics before and after the antibacterial treatment. This model uses a simplified description of the oral bacterial biofilm and performs as good as or even better than the multispecies model given in [14], making it a robust, quantitative model for study biofilm dynamics.

**Materials and Methods**

**Experiment Development**

*Biofilm specimen preparation and treatment*

Sterile hydroxyapatite (HA) discs (Clarkson Chromatography Products, Williamsport, PA) were used as the biofilm substrate. The HA discs were coated with bovine dermal type I collagen (10 μg/mL collagen in 0.012 N HCl in water) (Cohesion, Palo Alto, CA) as described in [16, 17]. Supragingival and subgingival plaque was collected from 2 adult volunteers, and suspended in brain heart infusion broth (BHI, Becton Dickinson, Sparks, MD, USA). For each plaque donor a separate batch of biofilms was grown. Coated HA disks were placed in the wells of a 24-well tissue culture plate containing 1.80 mL BHI, and 0.2 mL plaque suspension was added to each well. The disks were incubated in BHI under anaerobic conditions (Bactron300 Shellab Anaerobic Chamber; Sheldon Manufacturing, Inc., Cornelius, OR) at 37˚C for 3 weeks. Fresh medium was changed once per week.

After 3 weeks of anaerobic incubation in BHI broth, specimens were rinsed twice in 1 mL physiological saline for 1 minute and immersed in 1 mL 1% NaOCl, 0.2/0.4% iodine-potassium iodide (IPI), or 2% CHX for 10 minutes. One percent NaOCl was freshly prepared by diluting the 5.25% stock solution (The Clorox Company, Canada) in distilled water, 0.2/0.4% IPI was prepared by mixing 0.2 g iodine (Sigma Chemical Co, St Louis, MO) in 0.4 g potassium iodide (Sigma Chemical Co) and adding distilled water to a 100-mL volume, and 2% CHX was freshly prepared by diluting in sterile water from 20% stock solution (Sigma Chemical Co). Control specimens, after rinsing in saline, were exposed to 1 mL sterile commercialized water for 10 minutes.

*Examination with CLSM*

Samples for confocal laser scanning microscopy (CLSM) for viability staining were collected immediately after 1, 3, 8, 11 and 15 weeks after the exposure to the medicaments. Specimens tested with saline for corresponding time periods were used as controls. For all specimens, the fresh BHI broth was changed once a week.

Two biofilm discs were examined for each time period. The biofilms discs for CLSM imaging were rinsed in 0.85% physiological saline for 1 minutes to remove the culture broth. They were then stained with a bacterial viability stain (LIVE/DEAD *Bac*light Kit; Thermo Fisher Scientific, Waltham, MA) and scanned with CLSM as described previously [14-17]. Three-dimensional volume stacks were constructed with Imaris 7.2 software (Bitplane Inc, St Paul, MN), and the total volume of red (dead bacteria) and green (live bacteria) fluorescence was measured. The proportion of dead bacteria was calculated from the proportion of red fluorescence to the total of green and red fluorescence. The results were analyzed using Univariate ANOVA followed by post hoc analysis at a significance level of $P < 0.05$.

**Mathematical Model Formulation**

*Model assumptions*

The response of biofilms to antibacterial treatment in experiments reveals that the biofilm recovery process after antibacterial treatment is highly nonlinear and sensitive to antibacterial agents applied. A quantitative model would be useful to describe dynamics in the process and to identify influential factors (or biomarkers) dictating biofilm recovery after the treatment. In this model, we coarse-grain bacteria into two basic types: the live and the dead bacteria. Their volume fractions are denoted, respectively, as $L$ and $D$. Besides the bacteria, the volume fractions of the EPS and solvent are also non-negligible and are denoted by $E$ and $T$, respectively. There exist functional components in the biofilm that affect the growth/decay and dynamics of the biofilm. These include quorum sensing (QS) molecules, growth factor (GF) proteins and antibacterial agents in a minimal set. Since they normally occupy negligibly small volume fractions in the biofilm, their volume fractions are therefore ignored in this model for simplicity. Under these assumptions, we have
$$L + D + E + T = 1. \tag{1}$$

Previous experiments showed that biofilms would be mature at about the third week, and both the previous and recent experimental evidence showed that mature biofilms would undergo a long period to recover after antibacterial treatment. To model the nonlinear process of biofilm recovery, we introduce a functional component named growth factor in this model, to regulate cell proliferation. This is a proxy for perhaps a set of functional proteins. Identifying the growth factor or growth factors would be a challenging experimental endeavor in the future that we strongly recommend through this quantitative study. Here, we assume the concentration of nutrient during the process is constant for simplicity. The concentration of QS molecules, antibacterial agents and growth factors are denoted as *H*, *A* and *Q*, respectively.

*Dynamical equations of bacteria*

The birth and death of bacteria are dictated by several factors including bacterial proliferation, the natural cell death and the killing by antibacterial agents (*A*). The proliferation of live bacteria (*L*) is limited by nutrient and the carrying capacity (*L*$_{max}$) of the environment, and also affected by growth factors (*Q*). In this model, we assume the live bacteria proliferate following a logistic model with the growth rate regulated by the growth factor while the cell death is governed by the natural cause and the antibacterial killing. We use a causality diagram to show the mechanisms as follows:
$$L \xrightarrow{(c_2,Q,L)} 2L, \tag{2}$$

$$L \xrightarrow{(r_{bs}, c_3 \gamma A)} D, \tag{3}$$

where $c_2 \dfrac{Q^2}{Q^2 + k_q^2}$ is the proliferation rate of live cells, $r_{bs}$ is the natural death rate of live bacteria and $c_3 \gamma A$ the antibacterial killing rate. The reactive equation for live bacteria is then given by
$$\frac{dL}{dt} = c_2 \frac{Q^2}{Q^2 + k_q^2}\left(1 - \frac{L}{L_{max}}\right) L - r_{bs} L - c_3 \gamma A L, \tag{4}$$
where, $k_q$ is a constant in the Hill function and $\gamma$ is a relaxation parameter to represent the effect of spatial diffusion in the spatially homogeneous system, which is proposed in the Hinson model[18] as follows:
$$\gamma = \frac{1}{T + \frac{E}{D_{pr}}} \frac{2(T+E)}{2+(L+D)}, \tag{5}$$
where *D*$_{pr}$ is a parameter representing the relative diffusivity of EPS. In this approximation, spatial diffusion is effectively replaced in a uniform decay in space, i.e, the homogeneous spatial decay is used as a proxy for heterogeneous diffusion in space.

The natural and killing death of live bacteria contributes to the increase in the population of the dead bacteria. The dead bacteria (*D*) can also disintegrate to shed their surface-attached EPS. So, we assume the dead bacteria eventually disintegrate into EPS (*E*) and solvent (*T*):

$$D \xrightarrow{(r_{dp}, k_{13}, A)} E + T. \tag{6}$$

By assuming that the antibacterial agent can in fact slows down the degradation of dead cells, we arrive at the following reactive equation for the rate of change in dead bacterial volume fraction:

$$\frac{dD}{dt} = r_{bs} L + c_3 \gamma A L - r_{dp} \frac{k_{13}}{k_{13} + A} D, \tag{7}$$

where $r_{dp}$ is the break down rate of dead bacteria at the absence of antibacterial agents, $k_{13}$ is a constant. The first two terms are related to the death of live bacteria cells due to the natural cause and antibacterial effect, respectively, while the last term represents the degradation of dead bacteria into EPS and solvent components due to cell lysis as well as the antibacterial effect. Without being treated by antibacterial agents, the dead cell degrades with a constant rate $r_{dp}$. With the stress imposed by the antibacterial effect on the biofilm, we surmise the degradation of dead cells would be slowed down. This assumption on the degradation of dead cells is based on the assumption that antibacterial treatment disrupts the natural process of a cell cycle, delaying their disintegration into other biofilm components. This seemingly arbitrary assumption is in fact strongly supported by our model calibration in that the other assumptions on the decay rate would make the model less faithful to the experimental data. We therefore believe the assumption made is credible in this model.

*Reactive equation of EPS production*

The production of EPS ($E$) is affected by quorum-sensing molecules ($H$), live bacterial ($L$) and ultimately EPS saturation at its carrying capacity ($E_{max}$). The biochemical process can be described by the causality diagram:

$$L \xrightarrow{(c_5, H, E)} L + E, \tag{8}$$

where $c_5 \frac{H^2}{H^2 + k_9^2} \left(1 - \frac{E}{E_{max}}\right)$ is the EPS production rate. The reactive equation for the rate of change in EPS volume fraction is then given as follows:

$$\frac{dE}{dt} = \left(c_5 L \frac{H^2}{H^2 + k_9^2} + r_{dp} \frac{k_{13}}{k_{13} + A} D\right) \left(1 - \frac{E}{E_{max}}\right), \tag{9}$$

where the first part represents the gain of EPS due to live bacteria, the second part comes from the degradation of the dead bacteria, and $k_9$ is a constant in the Hill function for the QS concentration dominated production rate.

*Dynamical equations of the functional components*

The natural degradation of the antibacterial agents, loss in effectiveness, their diffusion and reaction with cells are considered in this quantitative model. The degradation process of antibacterial agents is modeled like this:

$$\frac{dA}{dt} = -c_8 A L - r_a A, \tag{10}$$

where $c_8$ is the killing rate of the bacteria by antibacterial agents and $r_a$ is the degradation of antibacterial agents in the solution.

The production of QS molecules ($H$) and growth factors ($Q$) are related to the volume fraction of live bacteria ($L$) while, in the meantime, can saturate at their carrying capacities $H_{max}$ and $Q_{max}$. For simplicity, we assume the increase in QS molecules ($H$) is also regulated by the growth factor ($Q$) in the form of a Hill function in this model. The mechanisms are summarized as follows:

$$L \xrightarrow{(c_a, Q, L, H)} L + H, \tag{11}$$

$$L \xrightarrow{(c_q, L, Q)} L + Q, \tag{12}$$

where $c_a, c_q$ are pre-factors for the QS molecules and growth factors, respectively. We propose the following reactive equations to quantify the mechanisms:

$$\frac{dH}{dt} = c_a \frac{Q^2}{Q^2 + k_q^2} L \left(1 - \frac{H}{H_{max}}\right), \tag{13}$$

$$\frac{dQ}{dt} = c_q L \left(1 - \frac{Q}{Q_{max}}\right), \tag{14}$$

where $H_{max}, Q_{max}$ are carrying capacities for QS molecules and growth factors, respectively.

*Summary of the governing equations*

The coupled dynamical equations of the biofilm in its dimensionless form are summarized as follows

$$\frac{dL}{dt} = c_2 \frac{Q^2}{Q^2 + k_q^2} (1 - \frac{L}{L_{max}}) L - r_{bs} L - c_3 \gamma AL, \tag{15}$$

$$\frac{dD}{dt} = r_{bs} L + c_3 \gamma AL - r_{dp} \frac{k_{13}}{k_{13} + A} D, \tag{16}$$

$$\frac{dE}{dt} = \left(c_5 L \frac{H^2}{H^2 + k_9^2} + r_{dp} \frac{k_{13}}{k_{13} + A} D\right) \left(1 - \frac{E}{E_{max}}\right), \tag{17}$$

$$\frac{dA}{dt} = -c_8 AL - r_a A, \tag{18}$$

$$\frac{dH}{dt} = c_a \frac{Q^2}{Q^2 + k_q^2} L \left(1 - \frac{H}{H_{max}}\right), \tag{19}$$

$$\frac{dQ}{dt} = c_q L \left(1 - \frac{Q}{Q_{max}}\right), \tag{20}$$

where

$$L + D + E + T = 1, \tag{21}$$

$$\gamma = \frac{1}{T + \frac{E}{D_{pr}}} \frac{2(T + E)}{2 + (L + D)}. \tag{22}$$

The causality relationships used to derive the dynamical equations are summarized in a computational graph in Fig 1. The nondimensionalization of the equations and variables is given in the appendix.

We solve this coupled dynamical system using ode45 solver in Matlab. Some model parameters are taken from the literature while others are calibrated against the experiments reported in this paper and in a previous publication [22].

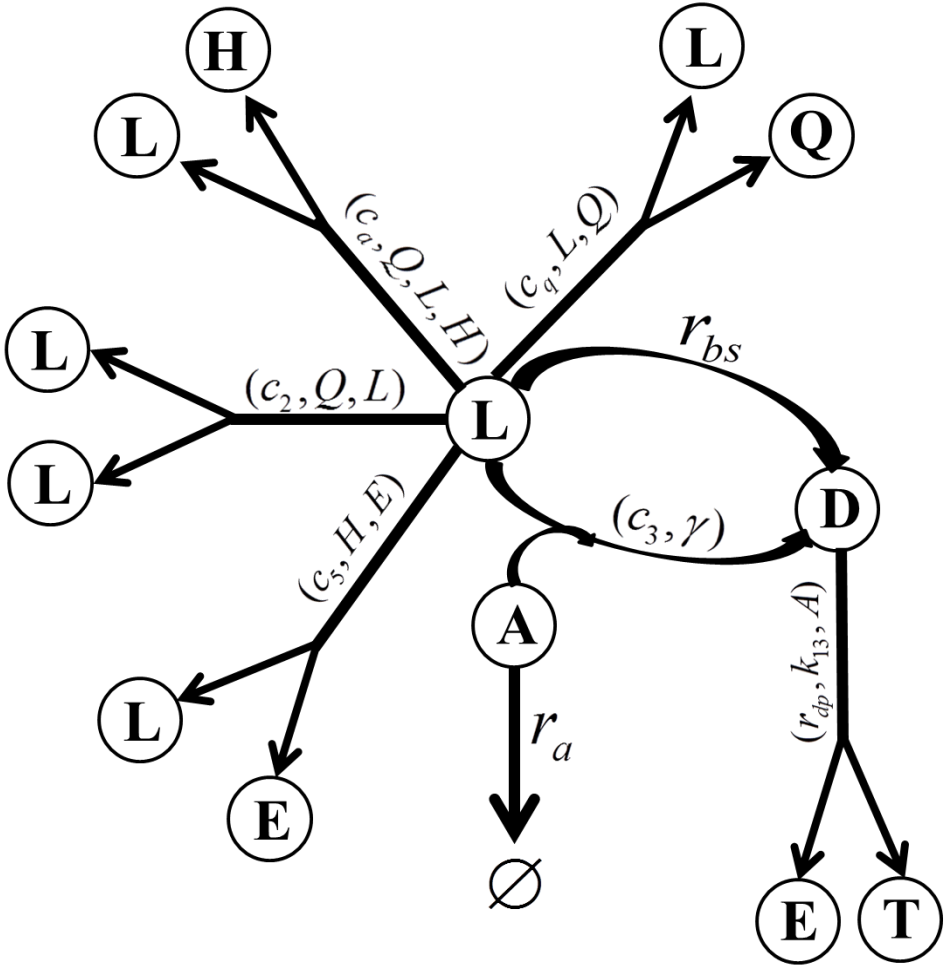

Figure 1. Computational graph of the mathematical model. *L*, *D*, *E*, *T*, *A*, *Q* and *H* are volume fractions or concentrations of live bacteria, dead bacteria, EPS, solvent, antibacterial agents, growth factors and QS molecules, respectively. ∅ means the degradation of the antibacterial agents. In this directed graph, causal relationships among the components are labelled with arrows and related rates.

**Calibration of Model Parameters**

The proposed model given in (15)~(22) is a mechanistic model based on a few fundamental mechanisms and experimentally informed assumptions. Several model parameters need to be calibrated against specific antibacterial agents and biofilm specimen from different donors. We use the following strategy to calibrate the parameters.

1. We first calibrate model parameters independent of antibacterial agents using the control group. Since there is no antibacterial agent in the control group, we set $A=0$. In addition, since the biofilm is fairly mature in the control group, we assume the volume fraction of live bacteria, dead bacteria and EPS in the control group have reached their steady state $L^{ss}$, $D^{ss}$, after three weeks. So we set

$$r_{bs}L^{ss} - r_{dp}D^{ss} = 0. \tag{23}$$

It follows that

$$r_{bs} = r_{dp}\frac{D^{ss}}{L^{ss}}. \tag{24}$$

Note that $r_{dp}/(r_{bs} + r_{dp}) = L^{ss}/(L^{ss}+D^{ss})$ and the live to total cell ratio is measured in the control group. So, we obtain the ratio of $r_{dp}$ and $r_{bs}$.

2. We then calibrate model parameters against the control group without the antibacterial effect. The parameters related to antibacterial agents are calibrated the last. After we obtain the ratio of $r_{dp}$ and $r_{bs}$, we use a bisection method to fit the parameters sequentially. We begin with a prescribed range of parameters resulting from the nondimensionalization and proceed until we get the best fit possible. Specifically, we set an upper limit for each parameter firstly. Then, we use the bisection method to search for the best fit of the model solution to the experimental data for the parameters one-by-one. Before finding the next parameter value using the bisection method, all the already obtained parameter values will be adjusted accordingly as the next parameter is fitted slightly. This method has been used to produce all the parameter values we use throughout this study.

We note that biofilms from different donors may be different. So, their responses to antibacterial treatment can also vary. Once the origin of the biofilm changes, we recalibrate the model parameters. As the result of the model calibration, we find that biofilm recoveries taken from different donors are similar, which indicates that there exist an invariant set of parameter values in the model that are insensitive to donors in biofilm recovery (see Table 2).

As a result, we classify the model parameters into three classes based on parameter calibration results: (I) donor-independent parameters; (II) donor-dependent parameters; (III) antibacterial agent specific parameters. Our study indicates that fluctuations of the parameters in group II on the donors are small. This result supports a single model for all donors' approach that we are taking in this study.

| Classes | I | II | III |
|---|---|---|---|
| Model parameters | $H_{max}$ | $c_2$ | $c_3$ |
| | $Q_{max}$ | $c_q$ | $c_8$ |
| | $c_5$ | $r_{bs}$ | $r_a$ |
| | $c_a$ | $r_{dp}$ | $k_{13}$ |
| | $c_9$ | $L_{max}$ | $D_{pr}$ |
| | $k_q$ | $E_{max}$ | |

**Table 1.** Classification of model parameters: (I) donor-independent parameters; (II) donor-dependent parameters; (III) antibacterial agent specific parameters

## Results

*Experimental results*

A total of 96 HA biofilm discs and 480 scanned areas were analyzed. The biofilms were treated for 10 mins by three antibacterial agents three weeks after they were taken from donors. Immediately after treatment, the viability profile of the biofilm population changed, demonstrating an increased number of dead cells (Fig 2). This occurred in all groups, but was more prominent in the biofilms treated with 1% NaOCl. NaOCl showed higher levels of bactericidal activity compared to 2% CHX and 0.2/0.4% IPI ($P < 0.05$; Fig 2). Cell death in the biofilms continued to intensifies for up to one week after exposure to all antibacterial solutions ($P < 0.001$). The viability of the bacterial population increased three weeks after treatment, although biofilms began to recover one week after treatment (Fig 2). Eight weeks after treatment, the proportion of viable bacteria almost reached that of the pre-treatment level in CHX groups, but was somewhat less in IPI and NaOCl treated groups (Fig 2). Eleven weeks after the treatment, bacterial viability in all groups returned to the pre-treatment level (expressed as percentages).

Figure 2. Recovery of bacteria in biofilms taken from donors after 10 min treatment using three antibacterial solutions. The effectiveness of bacterial killing ranks from NaCO1 to IPI to CHX.

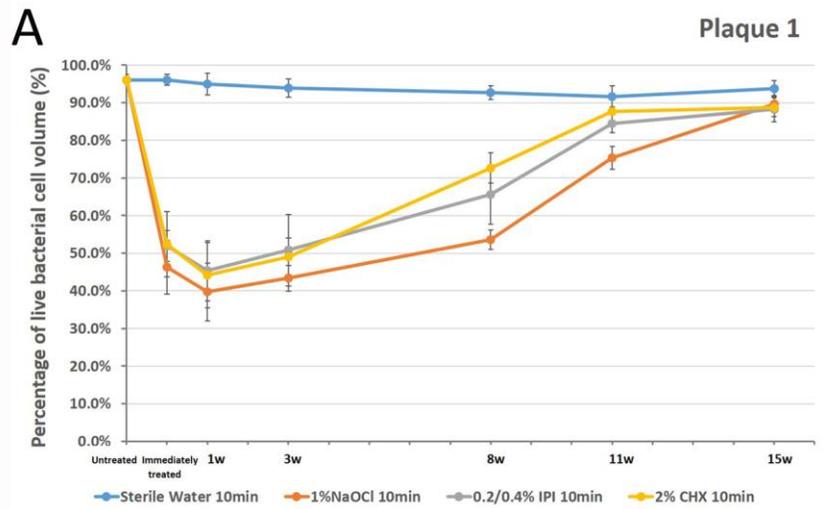
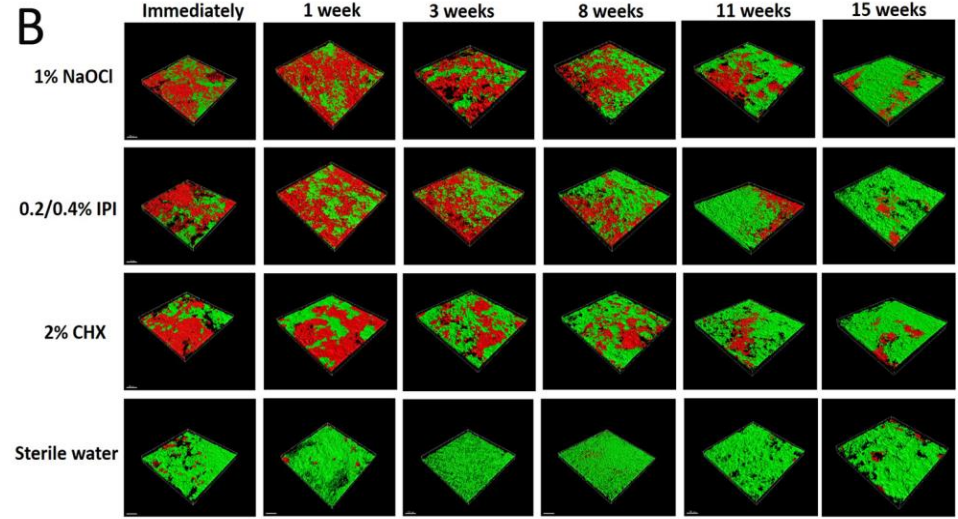
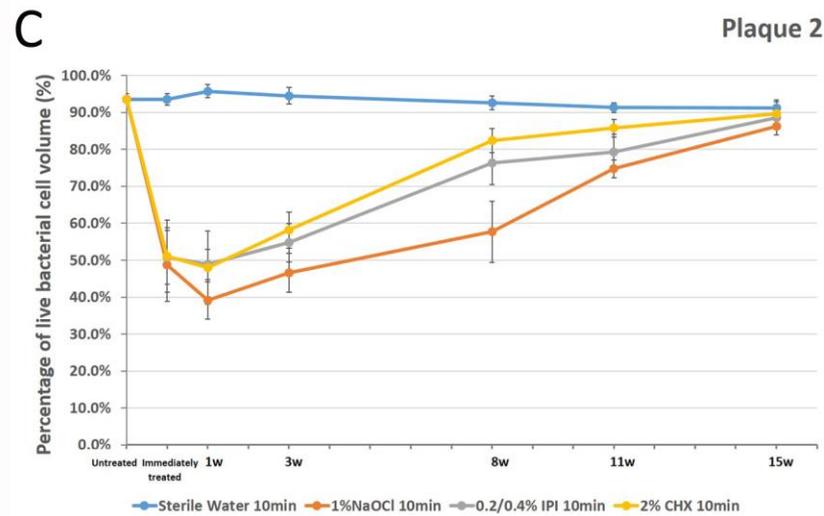
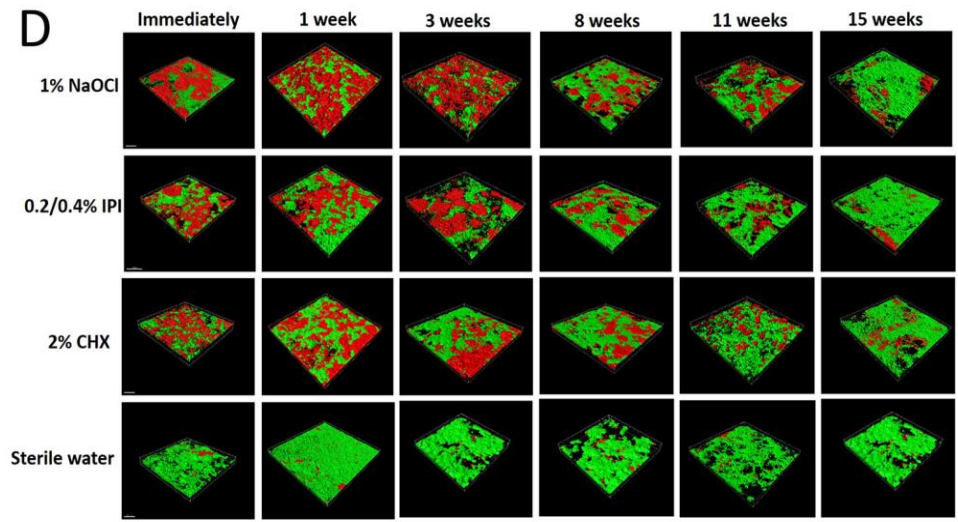

*Numerical Results*

The percentage of live bacteria dropped immediately after antibacterial treatment. Since the bacteria were treated at the end of the third week of biofilm growth, after taken from the donors, and the duration of treatment was 10 mins, we set the moment right after the treatment as time 0 in the dynamical simulations of the model. The percentage recovered to its pretreatment value at the end of the next 15 weeks. The numerical solution of the dynamical model all fall within the error bar of the experimental data, giving a reasonable prediction. As shown in Fig 3, the ability of NaOCl to kill bacteria is much stronger than that of CHX and IPI. So, the biofilm treated by NaOC1 recovers the most slowly.

In Fig 5(A&B), we plot the solution of the model prediction for donor 1. We notice that the decline of the live bacterial ratio initially after antibacterial treatment is to a large extent due to the rapid increase in the number of dead bacteria. When the number of dead bacteria starts to decline after a week, the ratio of live bacterial begins to increase as well.

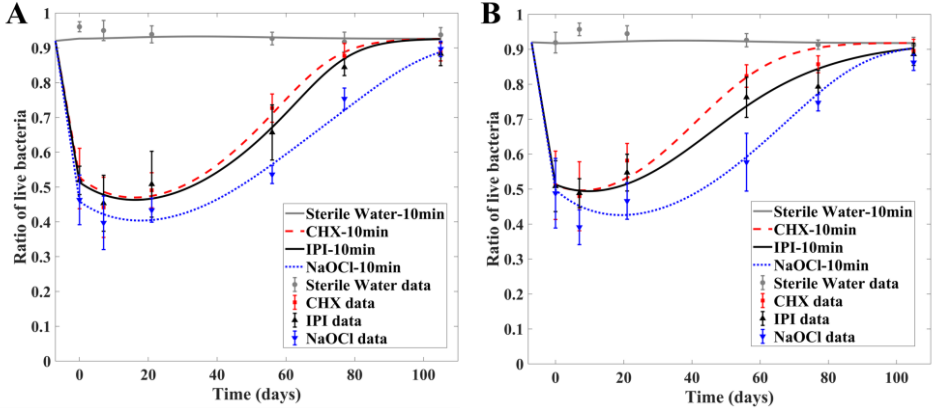

Figure 3. Time evolution of biofilm recovery for donor1 (A) and donor2 (B). The biofilms were taken from donor 1 and donor 2, respectively, and treated by three antibacterial solutions for 10 mins. At the time of antibacterial treatment, the biofilms had already grown for three weeks. The results indicate that NaOCl is the most effective antibacterial agent among the three and IPI shows slightly higher effectiveness than CHX for both donors. The ratio of live bacteria was calculated using $L/(L+D)$ and measured in the experiments, which are then used in parameter calibration. The smooth curves are the model predictions and the points with error bars are experimentally measured values.

| Symbol | Units | Donor 1 | | | Donor 2 | | |
|---|---|---|---|---|---|---|---|
| | | CHX | IPI | NaOCl | CHX | IPI | NaOCl |
| $H_{max}$ | kg/m$^3$ | 8.24e-3 | 8.24e-3 | 8.24e-3 | 8.24e-3 | 8.24e-3 | 8.24e-3 |
| $Q_{max}$ | kg/m$^3$ | 2.47e-3 | 2.47e-3 | 2.47e-3 | 2.47e-3 | 2.47e-3 | 2.47e-3 |
| $c_2$ | s$^{-1}$ | 3e-6 | 3e-6 | 3e-6 | 3e-6 | 3e-6 | 3e-6 |
| *$c_3$ | s$^{-1}$m$^3$/kg | 2.06e-1 | 1.5e-1 | 4.5e-1 | 1.76e-1 | 1.40e-1 | 5.81e-1 |
| $c_5$ | s$^{-1}$ | 9.9e-1 | 9.9e-1 | 9.9e-1 | 9.9e-1 | 9.9e-1 | 9.9e-1 |
| *$c_8$ | s$^{-1}$ | 2e-5 | 2e-5 | 9e-6 | 2e-5 | 1e-5 | 1e-5 |
| $c_a$ | kg/m$^3$ | 8.24e-9 | 8.24e-9 | 8.24e-9 | 8.24e-9 | 8.24e-9 | 8.24e-9 |
| $c_q$ | kg/m$^3$ | 1.65e-8 | 1.65e-8 | 1.65e-8 | 1.65e-8 | 1.65e-8 | 1.65e-8 |
| *$r_a$ | s$^{-1}$ | 1.8e-7 | 2e-7 | 3e-7 | 8e-8 | 2.4e-7 | 2.7e-7 |
| $r_{bs}$ | s$^{-1}$ | 1.6e-7 | 1.6e-7 | 1.6e-7 | 1.8e-7 | 1.8e-7 | 1.8e-7 |
| $r_{dp}$ | s$^{-1}$ | 2e-6 | 2e-6 | 2e-6 | 2e-6 | 2e-6 | 2e-6 |
| *$D_{pr}$ | | 1.6e-2 | 5e-2 | 6e-3 | 2.66e-2 | 7e-2 | 3.5e-3 |
| $k_9$ | kg/m$^3$ | 6.59e-3 | 6.59e-3 | 6.59e-3 | 6.59e-3 | 6.59e-3 | 6.59e-3 |
| *$k_{13}$ | kg/m$^3$ | 2.88e-8 | 1.48e-8 | 1.48e-8 | 1.24e-8 | 1.24e-8 | 1.48e-8 |
| $k_q$ | kg/m$^3$ | 2.47e-3 | 2.47e-3 | 2.47e-3 | 2.47e-3 | 2.47e-3 | 2.47e-3 |
| $L_{max}$ | | 8e-2 | 8e-2 | 8e-2 | 8e-2 | 8e-2 | 8e-2 |
| $E_{max}$ | | 1.2e-2 | 1.2e-2 | 1.2e-2 | 1.2e-2 | 1.2e-2 | 1.2e-2 |
| *The leaked agents | kg/m$^3$ | 1.24e-6 | 9.89e-7 | 1.3e-7 | 1.03e-7 | 1.25e-7 | 1.43e-7 |

**Table 2.** The model parameter values calibrated using the data from the control group and the treated group. The starred entries indicate different model parameters for different donors and the others are identical to all donors indicating they are insensitive to different donors. The characteristic time scale $t_0$=1s and the characteristic concentration scale $C$=8.24e-3kg/m$^3$ are used [19].

**Model predictions on responses of biofilms of various ages to antibacterial treatment**

We apply the mathematical model to another set of experimental data reported by Stojicic et al. in [6]. Here, we calibrate the model parameters to fit responses of various aged biofilms to the three antibacterial agents following the procedure alluded to earlier. Stojicic et al investigated the dynamic process of biofilms of various ages varying from the initial attachment of planktonic bacteria to a mature, structurally complex biofilm (0~8 weeks). The biofilms of various ages up to 8 weeks were treated by 2% CHX (Fig 4A), 0.2/0.4% IPI (Fig 4B) and 1% NaOCl (Fig 4C) for 1 min and 3 mins, respectively. Their work showed significant difference among the killing ratios of all tested antibacterial solutions in young (less than 2 weeks old) and mature (3 weeks or older) biofilms in all treatments. We use the quantitative model to simulate biofilm growth from the beginning to the moment right after the antibacterial treatment and report the post treatment result. The model fits very well to the experimental data following the parameter calibration procedure alluded to earlier. In addition, we notice that the recalibrated parameter values in the model do not differ much from the other set of parameter values calibrated on a data set of completely different biofilm experiments, indicating the model indeed captures the essence in biofilm growth dynamics. Both experiments and model predictions show that biofilms become mature after about three weeks (Fig 4). For mature biofilms, the consequence of antibacterial treatment becomes insensitive to the age of the biofilms. For younger biofilms the killing of live bacteria in treated biofilms is significant compared with the mature biofilms (see Fig 5 C&D). For mature biofilms, we observe that the EPS concentration is high at the time of treatment (Fig 5C). In the quantitative study, the dimensionless concentration of antibacterial agents we use to briefly treat the biofilms is chosen as 1 and initial volume fraction of live bacteria taken from the donors is chosen as 0.05. The experimental data are taken from the paper [6].

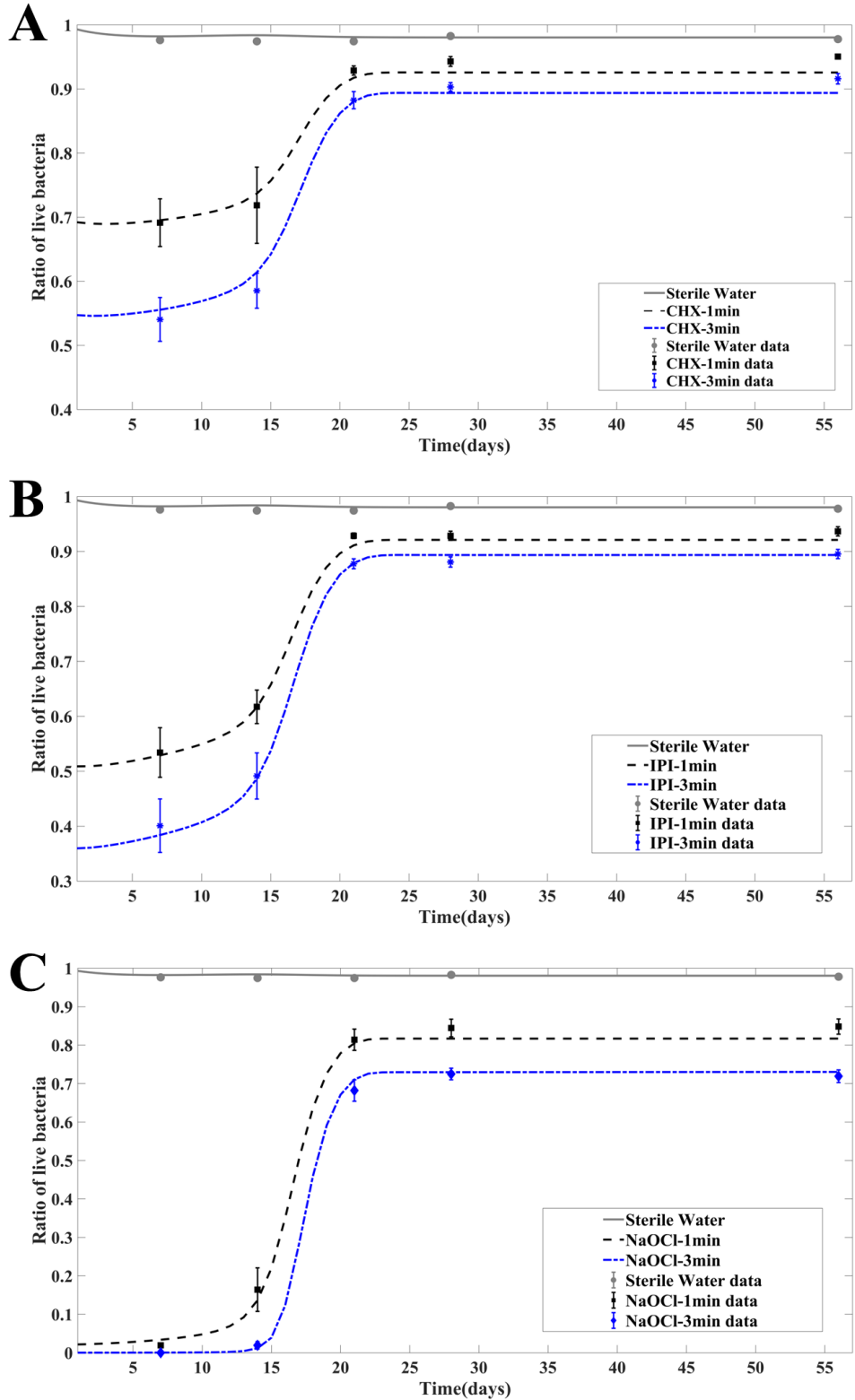

Figure 4 The responses of different aged biofilms to 1 min and 3 min CHX treatment (A), 1 min and 3 min IPI treatment (B) and 1 min and 3 min NaOCl treatment (C), respectively. The time axis here indicates the age of the biofilm. The results presented are the solutions of the quantitative model at different ages right after they are treated by antibacterial agents for 1 min and 3 mins, respectively. For example, the curve plotted at t=21 days are the solutions of the model at 21days+1

min (dashed) and 21days+3min (dotted) with the antibacterial agent applied to the biofilm at the 21st day for 1 min and 3 mins, respectively.

| Symbol | Units | CHX | IPI | NaOCl |
|---|---|---|---|---|
| $H_{max}$ | kg/m$^3$ | **8.24e-3** | **8.24e-3** | **8.24e-3** |
| $Q_{max}$ | kg/m$^3$ | **2.47e-3** | **2.47e-3** | **2.47e-3** |
| $c_2$ | s$^{-1}$ | **3e-4** | **3e-4** | **3e-4** |
| *$c_3$ | s$^{-1}$m$^3$/ kg | **1.20** | **2.18e-1** | **9.47e-1** |
| $c_5$ | s$^{-1}$ | **9.9e-1** | **9.9e-1** | **9.9e-1** |
| $c_8$ | s$^{-1}$ | **1e-2** | **1e-2** | **1e-1** |
| $c_a$ | kg/m$^3$ | **8.24e-9** | **8.24e-9** | **8.24e-9** |
| $c_q$ | kg/m$^3$ | **4.94e-9** | **4.94e-9** | **4.94e-9** |
| *$r_a$ | s$^{-1}$ | **1.3e-2** | **1.6e-2** | **4e-3** |
| $r_{bs}$ | s$^{-1}$ | **1.6e-7** | **1.6e-7** | **1.6e-7** |
| $r_{dp}$ | s$^{-1}$ | **5e-6** | **5e-6** | **5e-6** |
| *$D_{pr}$ | | **4.8e-2** | **2.66e-2** | **1.66e-3** |
| $k_9$ | kg/m$^3$ | **6.59e-3** | **6.59e-3** | **6.59e-3** |
| $k_{13}$ | kg/m$^3$ | **1.48e-8** | **1.48e-8** | **1.48e-8** |
| $k_q$ | kg/m$^3$ | **2.47e-3** | **2.47e-3** | **2.47e-3** |
| $L_{max}$ | | **1.2e-1** | **1.2e-1** | **1.2e-1** |
| $E_{max}$ | | **2.2e-1** | **2.2e-1** | **2.2e-1** |

**Table 3.** The model parameter values calibrated based on the response of different aged biofilms to antibacterial treatment. The starred entries indicate the parameters are sensitive to antibacterial agents in the model while the others are insensitive to antibacterial agents. The characteristic time scale $t_0$=1s and the characteristic concentration scale $C$=8.24e-3 kg/m$^3$ are used [19].

**Role of QS molecules and the growth factor in biofilm dynamics**

Since the recovery curves are the same qualitatively for different antibacterial agents and donors, we only show the time evolution of volume fractions of live bacteria (*L*), dead bacteria (*D*), EPS (*E*), the concentration of QS molecules (*H*) and the concentration of the growth factor (*Q*) in biofilms from donor 1 after 10 min CHX treatment as an example. Despite that the percentage of live bacteria continues to decrease as the result of a drastic increase in dead bacteria in the first week after the treatment (Fig 2&3), the volume fraction of live bacteria increases along with the growth factor right after the treatment (Fig 5 A&B). In this example, the results show that the EPS volume fraction reaches a plateau before the treatment even begins and is therefore less sensitive to the QS molecule and residual antibacterial agents after the treatment (Fig 5 A&B). As a comparison, we also plot the corresponding variables in the control group, labelled as untreated group. It shows clearly the live and dead bacteria eventually return to the level in the control group, while the QS molecule and the growth factor concentration never fully recover at the end of the simulation.

We also examine time evolution of the aforementioned biofilm components from the responses of different aged biofilms to 3 min CHX treatment (Fig 5 C&D). The residual volume fractions of *L*, *D* and *E* right after the antibacterial treatment vary little for mature biofilms older than 21 days right after treatment, which means that the effectiveness of bacterial killing by the antibacterial agents saturates with respect to the mature biofilms. Despite of the lack of efficiency in killing live bacteria in mature biofilms, the QS molecule and growth factor continues to grow to sustain the recovery of the biofilms. As a comparison, the solutions from the control group are also plotted in Fig 5 C&D. We observe that the 3 min antibacterial treatment does have a sizable impact on the live and dead bacteria for all aged biofilms and the reduction in live bacteria and increase in dead bacteria right after the 3 mins treatment for mature biofilms are indeed insensitive to the age for older biofilms (older than 21 days).

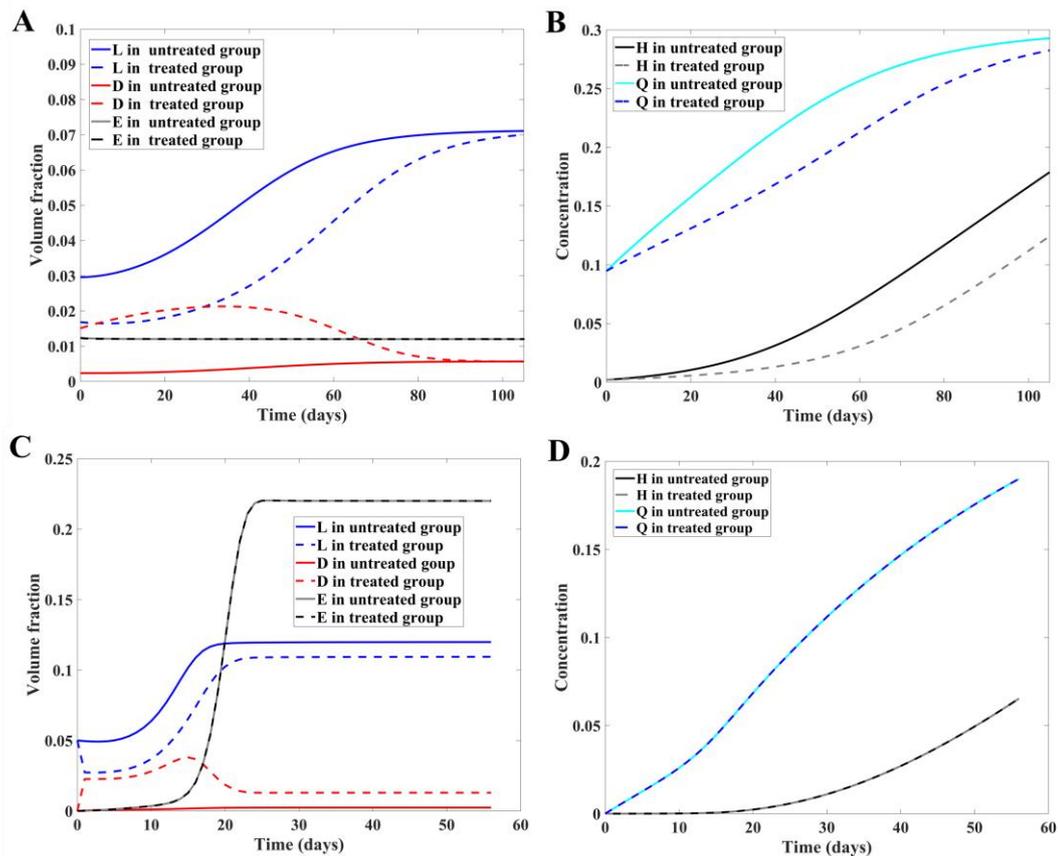

**Figure 5**. (A) and (B): Time evolution of volume fractions of live bacteria (*L*), dead bacteria (*D*), EPS (*E*), concentrations of QS molecules (*H*) and growth factor (*Q*) in the biofilm model without treatment and after 10 min CHX treatment for donor 1 in the experiment. The QS molecule and growth factor level in treated biofilms are lower than those in the control group, shown in (B). (C) and (D): The same set of selected variables predicted by the model without treatment and right after the 3 min CHX treatment with respect to biofilms of different age groups taken from [22]. The time axis in (C) and (D) indicates the age of the biofilm. All the quantities plotted on the curves are the solutions of the model right after the 3 min antibacterial treatment with the initiation condition taken from the control group (i.e., the untreated biofilm). The corresponding solutions from the control group are also plotted for comparison. In (C), volume fractions of live, dead bacteria and EPS saturate among the older biofilms for both the treated and the control group. The 3 min treatment induces negligibly small changes in concentrations of QS molecules and growth factor, shown in (D).

*Discussion*

Guided by the experiments, we have developed a quantitative model to describe biofilm recovery after antibacterial treatment. The multispecies oral biofilms grew out of the pooled plaque from two different donors showed nearly no difference in their susceptibility to all antibacterial agents, indicating a lack of variability in sensitivity from different sources of biofilms. The model can fit experimental observations well through a parameter calibration procedure. Although we coarse-grain the bacteria into only two gross types in the model for simplicity, the model can reproduce the recovery process very well for the given experimental data set. Study by Stojicic et al. [6] using multispecies biofilms grown on collagen-coated hydroxyapatite disks from different donors showed that biofilms from 6 different sources had a similar, time-dependent susceptibility pattern. Later, Yang et al[20]. using an infected dentin model also found that the multispecies biofilms from different donors showed similar susceptibility. The current study shows the same pattern of biofilm recovery after exposure to antibacterial agents in all biofilms taken from different donors. The present study further supports that the source and possible differences in the species composition of the multispecies biofilm have less impact on its susceptibility. To our knowledge, this is the first time the effect of the source of biofilms recovery is investigated. The results clearly indicate that the time needed for biofilms to recover is less dependent on the type of bacteria present in the biofilm.

NaOCl, CHX, and IPI were chosen for the study because they are common endodontic irrigants and have different mechanisms in antibacterial treatment. NaOCl attacks microbe's heat shock proteins causing the bacteria to form clumps and to die eventually [21]. CHX reacts with bacterial cellular membrane [22] or causes the precipitation of the cytoplasmic contents [23]. The mechanism of IPI involves multiple cellular effects by binding to proteins, nucleotides, and fatty acids[24]. CHX and IPI are less effective against biofilms than NaOCl. Post-antibiotic effect (PAE) is the continued suppression of bacterial growth after exposure of the bacteria to an antibacterial agent and removal of this agent from the environment [25-27]. Proposed mechanisms by which the PAE occurs include both nonlethal damage induced by the antibacterial agent and a limited persistence of the antibacterial agent at the bacterial binding site. Factors that affect the duration of the PAE include microorganism-antibacterial combination, duration of antibacterial agent's exposure, bacterial species, culture medium and experimental conditions. Interestingly, all irrigants have been shown to exhibit extended residual activity after treatment.

The new quantitative model developed in this study simplifies a previous model and even yields an improved fitting to the experimental data [14]. In this model, we group all unknown functional molecules/proteins into a single growth factor, which is a crucially important factor on bacterial growth and dynamics of QS molecules. Meanwhile, we assume dynamics of the growth factor is governed by the live bacteria and limited by its carrying capacity $Q_{max}$ in the model. The treatment of biofilms by the antibacterial agents does not show any significant effect on the growth factor, except that its growth right after the treatment is much slower than that at a later time. Interestingly, the tamed growth in the growth factor during the first week still fuels the *volume fraction* of live bacteria to increase rather than decrease despite that its absolute population is low. The population of dead bacteria increases initially after the treatment as expected. The overall ratio of live vs total bacteria predicted by the model matches the experimental data very well. The model fitting the recovery curves demonstrates the key role played by the growth factor on time evolution of the biofilms. Although QS molecules remain a steady increase after the treatment, they don't have any significant effect on the volume fraction of EPS in the biofilm overall. We have also experimented with other modeling approaches without the direct impact of the growth factor to bacterial growth without a success, which adds credibility to our current modeling approach.

The mathematical model is used to describe biofilms from various donors and being treated by various antibacterial solutions. Its sensitivity on model parameters is therefore very important. We combine the control group and treated biofilm data to develop a general parameter calibration protocol, by which we classify the model parameters into three classes: (I) donor-independent parameters: (II) donor-dependent parameters; (III) antibacterial agent specific parameters. We have demonstrated in this study that the model prediction fits quite well to the experimental data. So the parameters in class (I) have no relations with the donors and the antibacterial agents. They are constants in the model. Among the donor and antibacterial agent dependent parameters, our study shows that their fluctuations with respect to different donors and antibacterial agents are minimal. When the model is applied to a completely different data set from a previous experiment for studying biofilm dynamics right after antibacterial treatment of different aged biofilms, good results are obtained as well. For a mechanistic model for such a complex biological system, we believe that this model indeed captures the essence of the biofilm dynamics and can shed insight on investigating details in biofilm dynamics.

**Acknowledgments:**

Ya Shen's research is partially supported by a grant of Canada Foundation for Innovation (32623). Qi Wang's research is partially supported by grants of NSF DMS-1517347 award, NSFC awards #11571032, #91630207, NSAF-U1530401.

**Appendix:** Nondimensionalization

Notice that $L$, $D$, $E$, $T$ are dimensionless variables while $A$, $H$, $Q$ are not. We need to nondimensionalize the dimensional variables and equations in order to analyze and compute the equations. We denote the characteristic time scale as $t_0$, the common characteristic concentration as $C$ We choose $C$=8.24e-3 kg/m$^3$ [31] and $t_0$=1/s. We define the dimensionless parameters as follows:

$\hat{t} \triangleq \frac{t}{t_0}$, $\hat{c_2} \triangleq c_2 t_0$, $\widehat{k_q} \triangleq \frac{k_q}{C}$, $\widehat{r_{bs}} \triangleq r_{bs} t_0$, $\hat{c_3} \triangleq c_3 t_0 C$, $\widehat{r_{dp}} \triangleq r_{dp} t_0$, $\widehat{k_{13}} \triangleq \frac{k_{13}}{C}$, $\hat{c_5} \triangleq c_5 t_0$, $\widehat{k_9} \triangleq \frac{k_9}{C}$, $\hat{A} \triangleq \frac{A}{C}$, $\hat{c_8} \triangleq c_8 t_0$, $\hat{r_a} \triangleq r_a t_0$, $\hat{H} \triangleq \frac{H}{C}$, $\hat{Q} \triangleq \frac{Q}{C}$, $\widehat{H_{max}} \triangleq \frac{H_{max}}{C}$, $\widehat{Q_{max}} \triangleq \frac{Q_{max}}{C}$, $\hat{c_a} \triangleq \frac{c_a}{C} t_0$, $\widehat{k_q} \triangleq \frac{k_q}{C}$, and $\hat{c_q} \triangleq \frac{c_q}{C} t_0$.

Substituting these parameters into the equations (1)~(6), we obtain the dimensionless equations. If we drop the hat on the variables and the parameters, we recover the dimensionless equations in exactly the same form as the dimensional equations.

(1)

$$\frac{d\hat{Q}}{d\hat{t}} = \hat{c}_2 \frac{\hat{Q}^2}{\hat{Q}^2 + k_q^2}(1 - \frac{L}{L_{max}})L - \widehat{r_{bs}}L - \hat{c}_3\gamma\hat{A}L,$$

$$\frac{dD}{d\hat{t}} = \widehat{r_{bs}}L + \hat{c}_3\gamma\hat{A}L - \widehat{r_{dp}}\frac{\widehat{k_{13}}}{\widehat{k_{13}} + \hat{A}}D, \tag{2}$$

$$\frac{dE}{d\hat{t}} = (\hat{c}_5 L \frac{\hat{H}^2}{\hat{H}^2 + \widehat{k_9}^2} + \widehat{r_{dp}}\frac{\widehat{k_{13}}}{\widehat{k_{13}} + \hat{A}}D)\left(1 - \frac{E}{E_{max}}\right), \tag{3}$$

$$\frac{dA}{d\hat{t}} = -\hat{c}_8 \hat{A}L - \hat{r}_a \hat{A}, \tag{4}$$

$$\frac{dH}{d\hat{t}} = \hat{c}_a \frac{\hat{Q}^2}{\hat{Q}^2 + \widehat{k_q}^2} L \left(1 - \frac{\hat{H}}{\widehat{H_{max}}}\right), \tag{5}$$

$$\frac{d\hat{Q}}{d\hat{t}} = \hat{c}_q L \left(1 - \frac{\hat{Q}}{\widehat{Q_{max}}}\right), \tag{6}$$

where

$$L + D + E + T = 1, \tag{7}$$

$$\gamma = \frac{1}{T + \frac{E}{D_{pr}}}\frac{2(T+E)}{2+(L+D)}. \tag{8}$$

After fitting the parameters from the dimensionless model, we can get the corresponding value in the experiment by perform the transform:

$t \triangleq \hat{t}t_0$, $c_2 \triangleq \hat{c}_2/t_0$, $k_q \triangleq C\widehat{k_q}$, $r_{bs} \triangleq \widehat{r_{bs}}/t_0$, $c_3 \triangleq \hat{c}_3/(t_0 C)$, $r_{dp} \triangleq \widehat{r_{dp}}/t_0$, $k_{13} \triangleq C\widehat{k_{13}}$, $c_5 \triangleq \hat{c}_5/t_0$, $k_9 \triangleq C\widehat{k_9}$, $A \triangleq C\hat{A}$, $c_8 \triangleq \hat{c}_8/t_0$, $r_a \triangleq \hat{r}_a/t_0$, $H \triangleq \hat{H}C$, $Q \triangleq C\hat{Q}$, $H_{max} \triangleq C\widehat{H_{max}}$, $Q_{max} \triangleq C\widehat{Q_{max}}$, $c_a \triangleq \hat{c}_a C/t_0$, $k_q \triangleq C\widehat{k_q}$, and $c_q \triangleq C\hat{c}_q/t_0$.